\documentclass[letterpaper,twocolumn,showpacs,
prb
,superscriptaddress,email
,showkeys%
]{revtex4-1}

\usepackage[pdftex]{graphicx}
\usepackage{amssymb}
\usepackage{mathrsfs}
\usepackage{amsmath,amsfonts,latexsym}
\usepackage{array,tabularx,color}
\usepackage{dcolumn}  % Align table columns on decimal point
\usepackage[normalem]{ulem}
\usepackage{comment}
\usepackage{bm}
\usepackage{wasysym}
\usepackage{soul}

\newcommand{\RB}[1]{{\color{black}#1}}
\definecolor{DarkGreen}{rgb}{0.000000,0.392157,0.000000}
\newcommand{\RC}[1]{{\color{black}#1}}
\newcommand{\RD}[1]{{\color{black}#1}}

\definecolor{DarkOrchid}{rgb}{0.600000,0.196078,0.800000}
\newcommand{\textmoved}[1]{{\color{black}#1}}

\begin{document}

\title{Quantum reflections and the shunting of polariton condensate wave trains: implementation of a logic AND gate}

\author{C. Ant\'on}
\affiliation{Departamento de F\'isica de Materiales, Universidad Aut\'onoma de Madrid, Madrid 28049, Spain}
\affiliation{Instituto de Ciencia de Materiales ``Nicol\'as Cabrera'', Universidad Aut\'onoma de Madrid, Madrid 28049, Spain}

\author{T.C.H. Liew}
\affiliation{School of Physical and Mathematical Sciences, Nanyang Technological University, 637371, Singapore}

\author{J. Cuadra}
\affiliation{Departamento de F\'isica de Materiales, Universidad Aut\'onoma de Madrid, Madrid 28049, Spain}

\author{M.D. Mart\'in}
\affiliation{Departamento de F\'isica de Materiales, Universidad Aut\'onoma de Madrid, Madrid 28049, Spain}
\affiliation{Instituto de Ciencia de Materiales ``Nicol\'as Cabrera'', Universidad Aut\'onoma de Madrid, Madrid 28049, Spain}

\author{P.S. Eldridge}
\affiliation{FORTH-IESL, P.O. Box 1385, 71110 Heraklion, Crete, Greece}

\author{Z. Hatzopoulos}
\affiliation{FORTH-IESL, P.O. Box 1385, 71110 Heraklion, Crete, Greece}
\affiliation{Department of Physics, University of Crete, 71003 Heraklion, Crete, Greece}

\author{G. Stavrinidis}
\affiliation{FORTH-IESL, P.O. Box 1385, 71110 Heraklion, Crete, Greece}

\author{P.G. Savvidis}
\affiliation{FORTH-IESL, P.O. Box 1385, 71110 Heraklion, Crete, Greece}
\affiliation{Department of Materials Science and Technology, Univ. of Crete, 71003 Heraklion, Crete, Greece}

\author{L. Vi{\~n}a}
\email{luis.vina@uam.es}
\affiliation{Departamento de F\'isica de Materiales, Universidad Aut\'onoma de Madrid, Madrid 28049, Spain}
\affiliation{Instituto de Ciencia de Materiales ``Nicol\'as Cabrera'', Universidad Aut\'onoma de Madrid, Madrid 28049, Spain}
\affiliation{Instituto de F\'isica de la Materia Condensada, Universidad Aut\'onoma de Madrid, Madrid 28049, Spain}

\date{\today}

\begin{abstract}
We study the dynamics of polariton condensate wave trains that propagate along a quasi one-dimensional waveguide. Through the application of tuneable potential barriers the propagation can be reflected and multiple reflections used to confine and store a propagating state. Energy-relaxation processes allow the delayed relaxation into a long-living coherent ground state. Aside the potential routing of polariton condensate signals, the system forms an AND-type logic gate compatible with incoherent inputs.
\end{abstract}



\maketitle

%%%%%%%%%%%%%%%%%%%%%%%%%%%%%%%%%%%%%%%
%Introduction
%%%%%%%%%%%%%%%%%%%%%%%%%%%%%%%%%%%%%%%
\section{Introduction}
\label{sec:intro}

In the last years, research on exciton-polariton condensates in semiconductor microcavities has initiated a quest for novel technological applications, which could set the basis for a new generation of devices in the next decades. Different groups have reported on the usefulness of polaritons to realize new lasers,~\cite{Imamoglu1996,Christopoulos2007,bajoni2008,Schneider:2013fk} and there is a strong focus on the development of elements for optical information processing, including: switches,~\cite{Amo2010,Adrados:2011tw,De-Giorgi:2012fk} transistors,~\cite{Gao:2012tg,anton:261116,Ballarini:2013uq} diodes,~\cite{Khalifa:2008zr,Bajoni:2008ly,Tsintzos2008,Nguyen2013} amplifiers,~\cite{savvidis00:prl,saba01,Wertz:2012ee} and integrated circuits.~\cite{Espinosa-Ortega:2013fk,Sturm2013}

This research effort is motivated by the fusion of photon and exciton properties that appear in exciton-polariton systems, giving rise to a fast (picosecond scale) response time, a long (nanosecond scale) dephasing time\RB{~\cite{Savona97}} and strong nonlinearities. Different fundamental properties of out-of-equilibrium polariton Bose-Einstein condensates have been profusely investigated, including coherence,~\cite{kasprzak06:nature,Balili2007,lai07a,Roumpos2012,Spano:2013bh} robust propagation of polariton wave trains,~\cite{Amo2009} frictionless flow,~\cite{Amo2009a} persistent quantized superfluid rotation,~\cite{Sanvitto2010,Marchetti2010} and solitary waves.~\cite{Amo:2011qf,Sich2012,Grosso2011} These phenomena, together with the capability of polariton condensates to couple to external light sources, form an ideal basis for the construction of optical information processing devices.

However, the development of optical information processing is heavily dependent on integrability with existing (i.e., electrical) technologies, which may well require designs using incoherent or non-resonant carrier injection methods. For this reason we focus on non-resonantly excited polariton condensates, in contrast to previous studies aimed at developing functional logical switches based on polaritons.~\cite{Amo2010,Adrados:2011tw,De-Giorgi:2012fk} In our previous work~\cite{anton:261116} we have found that repulsive interactions between polaritons and hot excitons in the system, which are inevitably excited with a non-resonant scheme, allow the acceleration of polariton condensates. This is consistent with earlier experiments by Wertz, et al.,~\cite{Wertz:2012ee} and measurements under continuous wave excitation.~\cite{Wertz2010,Gao:2012tg} In this work, we introduce a time-dependent control of propagating polariton condensates using optically-induced tuneable repulsive potential barriers. These barriers block propagation, causing reflection of an incident polariton condensate wave train. The presence of multiple barriers allows multiple reflections and the re-routing of the wave train into a confinement region. Unlike previous works on the gating of polariton condensates, the confined polaritons are still propagating and in excited states.

Considering the signal represented by propagating polariton condensates, we also demonstrate the ability of the system to function as an \emph{AND} type logic gate. When the propagating condensate is strongly confined, the energy-relaxation present in the system causes the formation of a long-living coherent ground-state. Taking this as the output state, \RB{its} formation requires both the initial injection beam and the presence of the optically-induced barrier. Our approach using only non-resonant excitation anticipates the implementation of new ultrafast, non-linear switches based on the electrical injection of polariton condensates.

%%%%%%%%%%%%%%%%%%%%%%%%%%%%%%%%%%%%%%%
%Sample and experimental setup
%%%%%%%%%%%%%%%%%%%%%%%%%%%%%%%%%%%%%%%
\section{Sample and experimental setup}
\label{sec:sample}

The sample used in the experiments is a high-quality $5\lambda/2$ AlGaAs-based microcavity containing 12 embedded quantum wells, giving a Rabi splitting of $\Omega_R = 9~$ meV \RB{(further information about this sample is given in Ref. \onlinecite{tsotsis12})}. Reactive ion etching has been applied to sculpt ridges with dimensions $20 \times 300~\mu$m$^2$. We select a region on the sample where the detuning between the bare exciton ($E_X$) and bare cavity mode ($E_C$) in the ridge corresponds to resonance, i.e. $\delta = E_C - E_X \backsim$ 0. We keep the sample at 10 K in a cold-finger cryostat and excite it with 2 ps-long light pulses from a Ti:Al$_2$O$_3$ laser, tuned to the bare exciton energy level (1.545 eV). Two independent, twin beams, dubbed \emph{A} and \emph{B}, are split from the laser: their intensities, spatial positions and relative time delay ($\Delta t=t_B-t_A$) can be independently adjusted. We have determined that the power threshold to produce polariton condensates is $P_{th}=4.4$ mW. \textmoved{The two laser beams used in the experiments are focused on the sample through a high numerical-aperture (0.6) lens, to form 10-$\varnothing$~$\mu$m spots. The same lens is used to collect and direct the emission towards a 0.5 m spectrometer coupled to a streak camera, working in synchroscan mode, obtaining energy-, time- and space-resolved images, with resolutions of 0.4 meV, 10 ps and 1~$\mu$m, respectively. Every picture is the result of an average over millions of shots (laser repetition rate 82 MHz and integration time 1.1 s).} The photoluminescence can be simultaneously resolved in the near- as well as in the far-field. \textmoved{The momentum space is simply accessed by imaging the Fourier plane of the lens used to collect the emission, taking advantage of the direct relation between the angle of emission and the in-plane momentum of polaritons.~\cite{richard05}}

\begin{figure}[!htbp]
\setlength{\abovecaptionskip}{-5pt}
\setlength{\belowcaptionskip}{-2pt}
\begin{center}
\includegraphics[width=1.0\linewidth,angle=0]{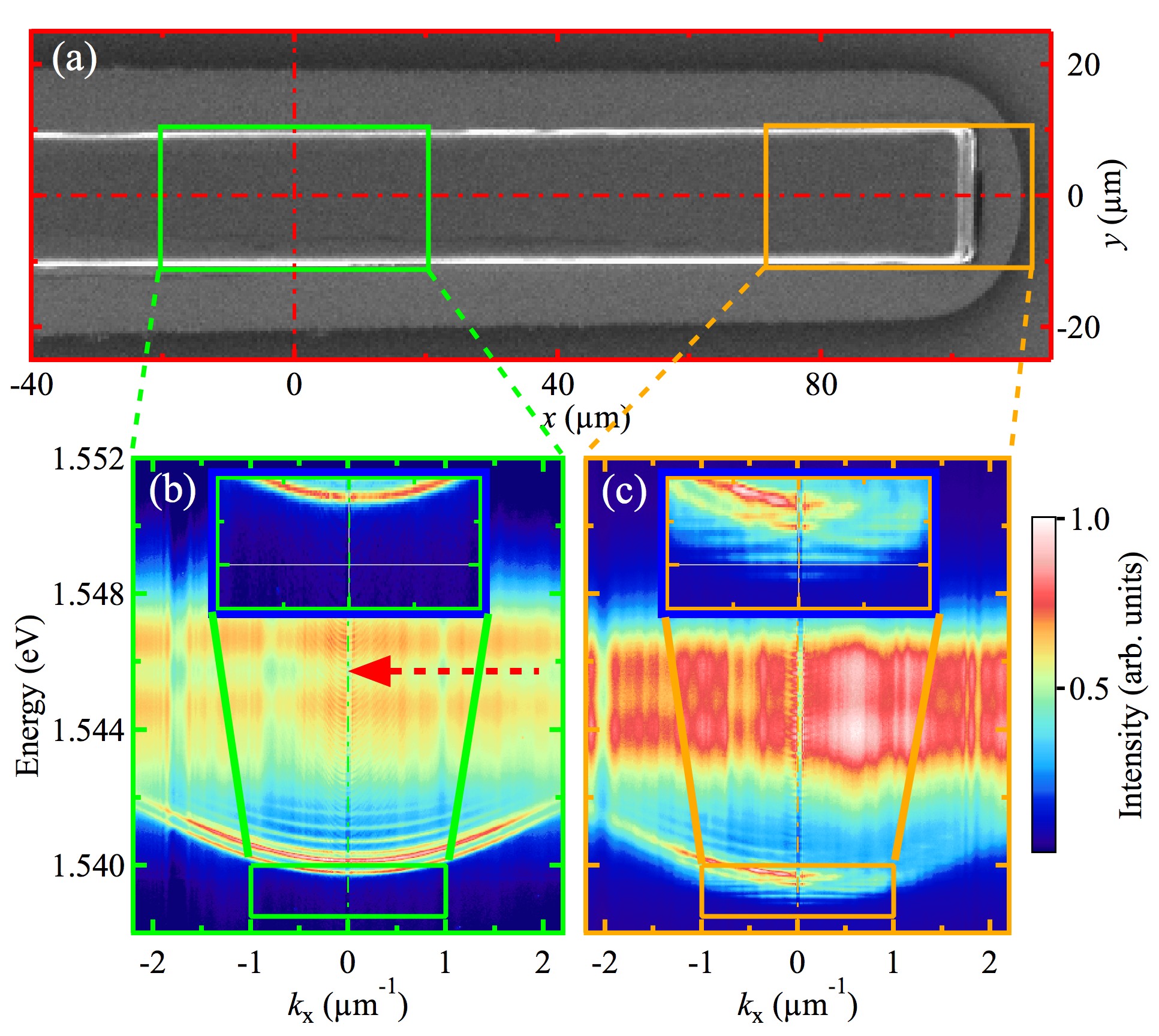}
\end{center}

\caption{\textmoved{(Color online) (a) Scanning electron microscopy image of a 20-$\mu$m wide ridge, including the filtered areas where time-integrated dispersion relations were measured under weak, non-resonant (1.612 eV), pulsed excitation: (b) far from its borders; (c) at the right ridge's border. The insets in panels (b) and (c) compare the dispersion relation quantization of polaritons at low energies. The dashed arrow indicates the energy of the excitation laser. The intensity is coded in a linear, normalized, false color scale.}}
\label{fig:disp_rel}
\end{figure}

\textmoved{In order to give a comprehensive characterization of the sample, we present now the dispersion relations of polaritons in relevant, different ridge's regions used in the experiments, obtained under non-resonant (1.612 eV), weak excitation conditions ($\sim$20 $\mu$W). Figure~\ref{fig:disp_rel} (a) shows a scanning electron microscopy image of the 20-$\mu$m wide ridge. Two rectangles centered at $x=$ 0 and 90~$\mu$m mark the spatial regions from where the real-space emission was collected in order to obtain the corresponding dispersion relations. The different sub-branches, originating from the quasi 1-D confinement in the $y$ direction of the ridge, are clearly resolved at $x=0$ in Fig.~\ref{fig:disp_rel}(b). The two bands centered at 1.545 and 1.547 eV correspond to the emission from bare exciton levels. The experiments described below were carried \RB{out by} exciting at the energy indicated by the dashed arrow. The dispersion obtained close to the ridge's border, Fig.~\ref{fig:disp_rel} (c), reveals the lack of translational invariance of polaritons at this position, evidenced by the formation of additional, non-dispersive low energy states at $\backsim$1.5392, 1.5394, 1.5395 and 1.5396 eV. Since the dispersion relation is obtained at the right ridge's border, only states in the dispersive branches with negative $k_x$ are occupied by left propagating polaritons.}

%%%%%%%%%%%%%%%%%%%%%%%%%%%%%%%%%%%%%%%
%Experimental results
%%%%%%%%%%%%%%%%%%%%%%%%%%%%%%%%%%%%%%%
\section{Experimental results and discussion}
\label{sec:exp}

\subsection{Optimal conditions for \emph{AND} gating}
\label{subsec:and}

\begin{figure}[!htbp]
\setlength{\abovecaptionskip}{-5pt}
\setlength{\belowcaptionskip}{-2pt}
\begin{center}
\includegraphics[trim=0.5cm 0.20cm 0.2cm 0.3cm, clip=true,width=1.0\linewidth,angle=0]{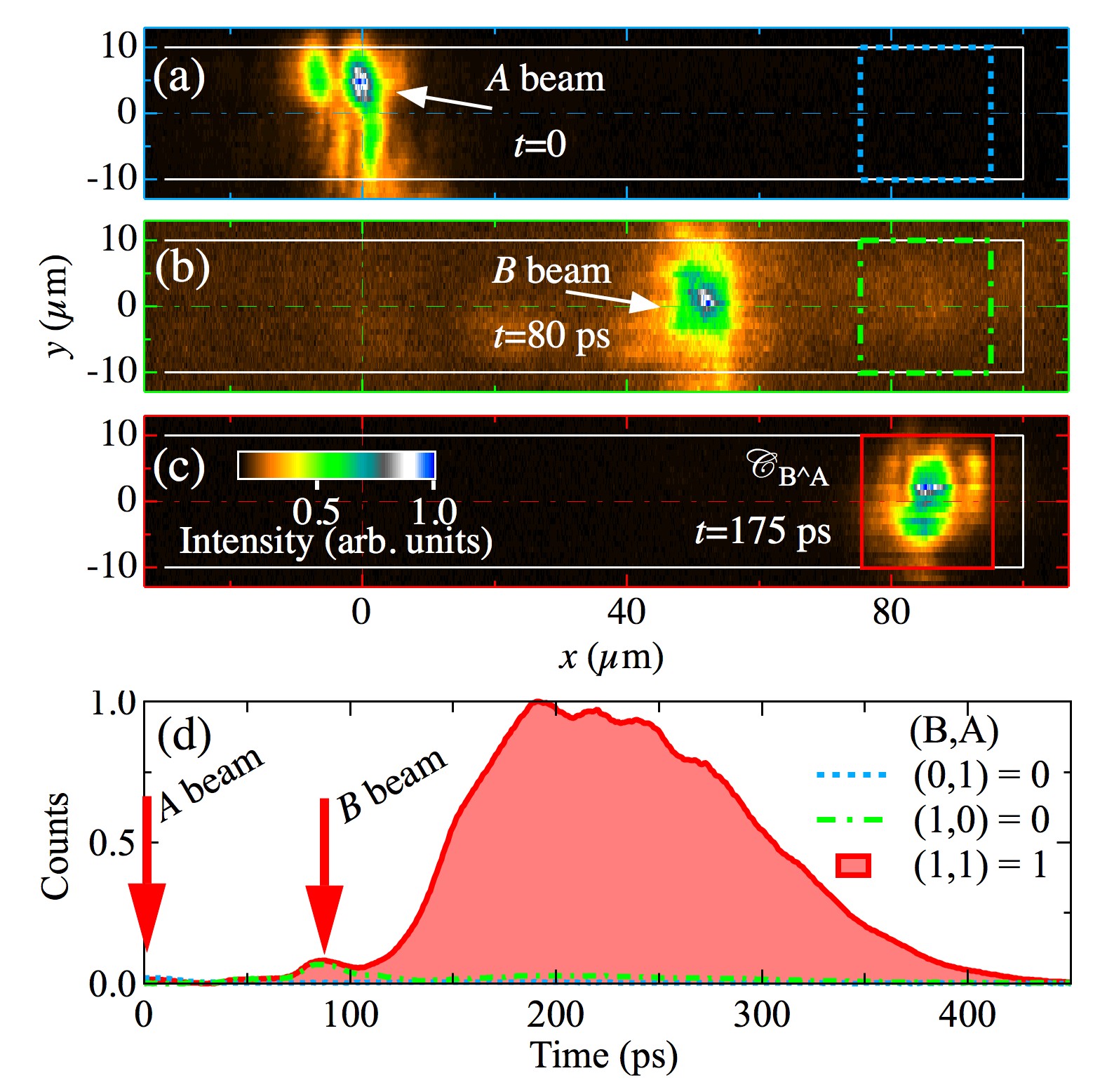}
\end{center}
\caption{(Color online) Real space intensity emission of the ridge at 1.539 eV: (a) scattered reflection of \emph{A} pulse impinging on the sample at $t=0$ and at $x = 0$; (b) scattered reflection of a second \emph{B} pulse, with a temporal delay of $\Delta t =$ 80 ps, at $x=50$~$\mu$m; (c) output polariton emission, $\mathscr{C}_{B \wedge A}$, (at 175 ps after the arrival of the \emph{A} pulse) close to the ridge's border, $x=85$~$\mu$m. \RC{The dotted, dot-dashed and full line boxes enclose the region of the output signal 0, 0 and 1, respectively.} The intensity is coded in a normalized, linear, false color scale. (d) \RC{Corresponding time evolution of the spatially integrated intensity from the boxes described before, the three traces are background-noise subtracted and normalized to the maximum of the $(B,A)=(1,1)$ trace. The vertical arrows mark the arrival of the A and B beams at 0 and 80 ps, respectively.}}
\label{fig:setup}
\end{figure}

To achieve the \emph{AND} gating operation of the device a proper choice of (i) the \emph{A} and \emph{B} beam locations, (ii) their power and (iii) the delay, $\Delta t$, between them must be made. Concerning the power, for the experiments described in this Subsection we have used $P_A=7\times P_{th}$ and $P_B=3\times P_{th}$, which are appropriate to obtain the \emph{AND} response of the device. Although we have used a non-resonant excitation, the power requirements remain comparable to previous microcavity switch designs based on hysteresis control ($>30$ mW)~\cite{De-Giorgi:2012fk} or resonant blueshift ($\sim5$ mW).~\cite{Ballarini:2013uq} As for the positioning and $\Delta t$, the \emph{A} beam is located $\backsim$100 $\mu$m away from the right ridge's border, Fig.~\ref{fig:setup} (a), the \emph{A} and \emph{B} beams are spatially separated by $\sim$50~$\mu$m and $\Delta t\approx$ 80 ps, Fig.~\ref{fig:setup} (b). 

As described in Ref. \onlinecite{Wertz2010}, the photo-generated excitons within the excitation area of a given beam, \emph{A} (\emph{B}), create a repulsive potential barrier; in our case at $\{x,t\}=\{0, 0\}$ ($\{x,t\}=\{$50~$\mu$m, 80 ps\}), labeled as $V_A$ ($V_B$). As discussed below, $V_A$ and $V_B$ determine the dynamics of the propagating polaritons in the ridge. Under the experimental conditions $(B,A)=(1,1)$ (indicating that both beams excite the sample) we obtain at long times a quasi-static, trapped polariton condensate, dubbed as $\mathscr{C}_{B \wedge A}$, Fig.~\ref{fig:setup} (c), which constitutes the [ON] state ($B \wedge A = 1$) of the device. The other states, given by $(B,A)=(0,1)$ and $(1,0)$, correspond to the [OFF] states ($B\wedge A=0$). Figure~\ref{fig:setup}(d) shows the evolution of the emission, spatially-integrated in the enclosed areas by the squares in Figs.~\ref{fig:setup} (a-c). The emission from $\mathscr{C}_{B \wedge A}$ (filled area) displays a fast rise time after the arrival of the \emph{B} pulse, with its maximum obtained at $\sim200$ ps; this is followed by a decay, with a weakly oscillating behavior. The corresponding emission, from the same enclosed region, in the \emph{A}-only (\emph{B}-only) configuration is negligible in this scale, see dotted (dot-dashed) trace, verifying the \emph{AND} operation of the device.

\begin{figure*}[!htbp]
\setlength{\abovecaptionskip}{-5pt}
\setlength{\belowcaptionskip}{-2pt}
\begin{center}
\includegraphics[trim=0.5cm 0.2cm 0.1cm 0.3cm, clip=true,width=1.0\linewidth,angle=0]{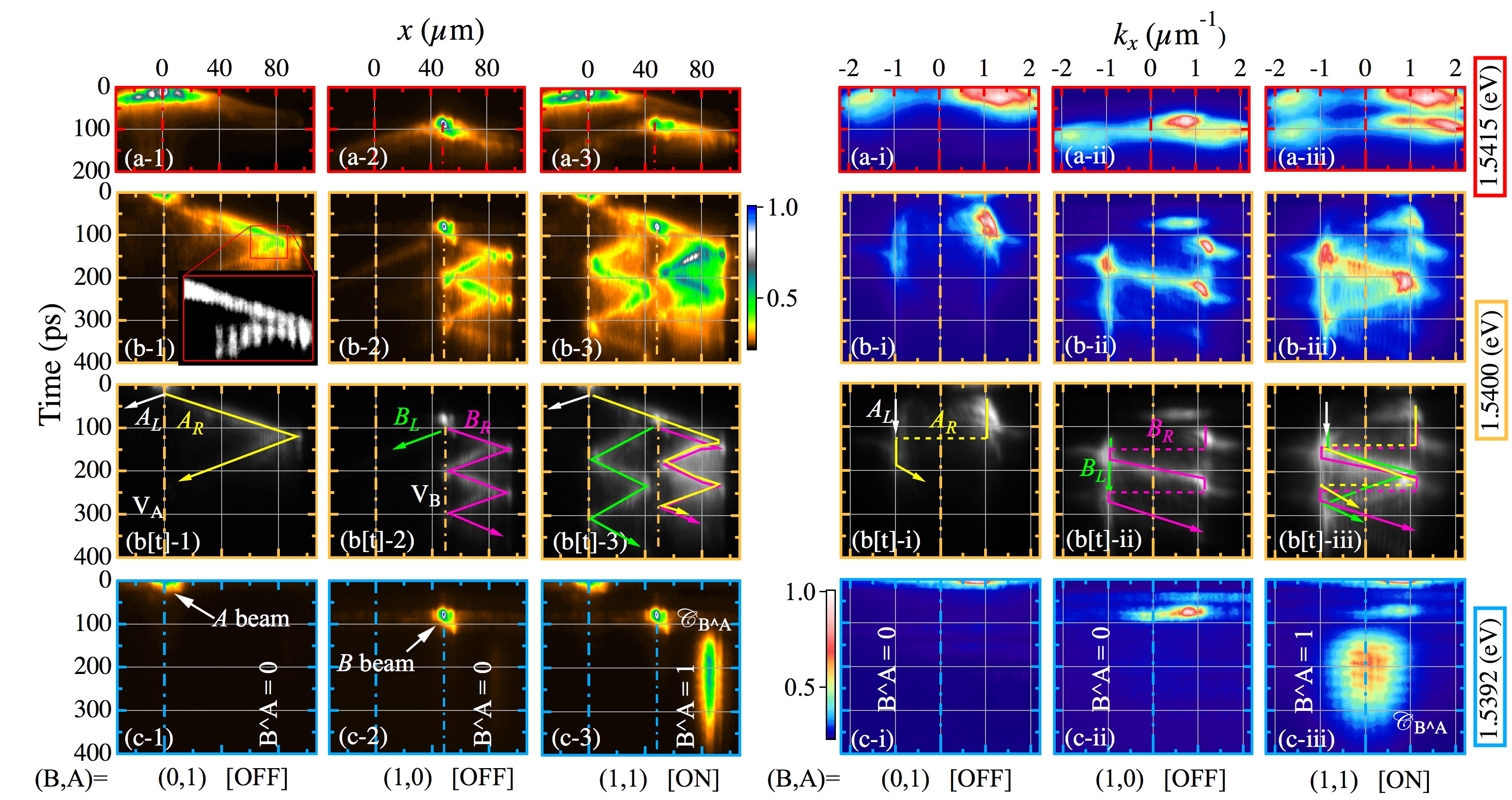}
\end{center}
\caption{(Color online) Real/Momentum-space dynamics of the polariton emission in the left/right group of panels, exciting with three logic address inputs, $(B, A)$ (see lower labels): $(0,1)$ \emph{A}-only [columns (1,i)]; $(1,0)$ \emph{B}-only [columns (2,ii)] and $(1,1)$ $A+B$ beams [columns (3,iii)], at different detection energies: 1.5415 eV (a), 1.5400 eV (b,b[t]) and 1.5392 eV (c). The inset in (b-1) shows the detail of the elastic reflection of $A_R$ against the ridge's border in the framed area. The trajectories of the polariton condensates at 1.540 eV are sketched by colored arrows in row b[t], on a background, false, grey color scale for the corresponding intensities in row (b). The intensities in real- and momentum-space are coded in logarithmic, normalized, false color scales, shown next to the panels. The movies showing the $x-y$ and $k_x-k_y$ polariton dynamics, compiled in panels (a-i), (b's) and (c-3)/(c-iii), are available as Supplemental Material.~\cite{SuppAnd}}
\label{fig:RK}
\end{figure*}

The truth table of the device in real- (left group of panels) and momentum-space (right group of panels) is summarized in Fig.~\ref{fig:RK}. The emission along the perpendicular coordinate, $y$/$k_y$, has been integrated. \RB{The polariton dynamics is shown at three-selected different energies, in the rows (a) 1.5415, (b) 1.5400 and (c) 1.5392 eV, since a full understanding of the device operation is only obtained when a spectroscopic analysis of its emission is performed.} For the sake of clarity, we have included an additional row, labeled (b[t]), where the trajectories in real- and momentum-space are sketched with colored arrows. The \emph{A} (\emph{B}) beam creates two, initially left/right propagating condensates along the $x$-axis, named $A_L$/$A_R$ ($B_L$/$B_R$).

We start describing the dynamics of the system under only one beam excitation. Columns (1) and (i) show, for the different detection energies, the configuration $(B,A)=(0,1)$, where the output address level reads zero, [OFF]. Figure~\ref{fig:RK} (a-1) displays hot polaritons propagating rapidly away from the \emph{A} excitation area (the large intensities at very short times arise partially from scattered laser light), and subsequently decaying into lower energy states, Fig.~\ref{fig:RK} (b-1), where an elastic reflection of the $A_R$ polariton condensate at the ridge's border is clearly observed at $\sim$125 ps. $A_R$ reaches the hill of $V_A$ at $\{x$[$\mu$m],$t$[ps]$\} \approx \{0, 225\}$, as depicted in Fig.~\ref{fig:RK} (b[t]-1) (at this instant the emission is very weak). It is remarkable that interference fringes exist in the polariton emission after the elastic reflection of $A_R$ at the border of the ridge, see the zoomed inset in Fig.~\ref{fig:RK} (b-1). They evidence the system coherence, even after the energy loss processes experienced by the original condensate created close to \emph{A}. Figure \ref{fig:RK} (c-1) shows only scattered light arising from the \emph{A} pulse at $\{x,t\}=$\{0,0\}; the absence of any emission \RB{establishes} the [OFF] state.

The corresponding momentum space dynamics is compiled in the fourth column, (i), of Fig.~\ref{fig:RK}. Hot polaritons at 1.5415 eV show a quasi homogeneous distribution of momenta in a \textbf{k}-space ring (see Supplemental Movie 1); these polaritons rapidly decay in energy, Fig.~\ref{fig:RK} (b-i) (1.540 eV): in the time interval 50 to 100 ps, $A_L$ and $A_R$ propagate at $k_x=\pm1.1$~$\mu$m$^{-1}$, as sketched in Fig.~\ref{fig:RK} (b[t]-i). The elastic reflection of $A_R$ at $\sim$125 ps is evidenced by the jump observed in \textbf{k}-space from $+1.1$ to $-1.1$~$\mu$m$^{-1}$ (horizontal segment of the dashed yellow arrow in (b[t]-i)). Figure \ref{fig:RK} (c-i) (1.5392 eV) demonstrates the [OFF] state, where only scattered light by \emph{A} is present.

We discuss now the second configuration where the output address level reads zero under \emph{B}-only excitation, $(B,A)=(1,0)$, shown in columns (2) and (ii), for the three detection energies. Figure \ref{fig:RK} (a-2) displays hot polaritons created at $t=$80 ps, propagating rapidly away from the \emph{B} area and subsequently decaying to lower energy states, Fig.~\ref{fig:RK} (b-2), in a similar fashion to what has been described before in the $(B,A)=(0,1)$ configuration, but with a more conspicuous $B_R$ trajectory due to the vicinity of $V_B$ and the ridge's border. $B_L$ moves at a constant speed of $v_x=-0.74$~$\mu$m/ps and the several reflections of $B_R$ take place at the positions sketched by the trajectory line, \RB{shown} in Fig.~\ref{fig:RK} (b[t]-2). At each reflection against $V_B$, where a reservoir of excitons exist, an amplification of $B_R$ is observed, as previously reported in Ref. \onlinecite{Wertz:2012ee}. Figure \ref{fig:RK} (c-2) shows information concerning only the scattered light by the \emph{B} pulse at $\{x$[$\mu$m],$t$[ps]$\}=\{50, 80 \}$.
%$\{x$[$\mu$m],$t$[ps]$\}$: \{95, 150\}-1$^{st}$, \{50, 200\}-2$^{nd}$, \{95, 250\}-3$^{rd}$ and \{50, 305\}-4$^{th}$
The corresponding momentum-space dynamics for the configuration $(B,A)=(1,0)$ is compiled in the fifth column, (ii), of Fig.~\ref{fig:RK}. The momenta of hot polaritons at 1.5415 eV show a similar behavior to that discussed previously for the $(B,A)=(0,1)$ case, see Fig.~\ref{fig:RK} (a-i). The evolution of the emission in \textbf{k}-space, appearing in Fig.~\ref{fig:RK} (b-ii), evidences the four reflections of $B_R$ in the same instants as those in Fig.~\ref{fig:RK} (b-2). The 1$^{st}$ and 3$^{rd}$ reflections show a sudden reversal in the propagation direction during an elastic scattering process (horizontal dashed segments of the arrow in b[t]-ii). However, the 2$^{nd}$ and 4$^{th}$ ones correspond to a continuous trade off between potential and kinetic energy (slanted segments of the arrow in b[t]-ii): $B_R$ propagates towards $V_B$ decelerating until it comes to a halt, and then it flows back towards the edge of the ridge; the continuous conversion of kinetic into potential energy are observed as continuous lines in the $k_x$ vs. time diagram from $-1.1$ to $+1.1$~$\mu$m$^{-1}$. Figure \ref{fig:RK} (c-ii), shown for completeness, displays only the scattered light by \emph{B}.

The [ON] state of the device is demonstrated in columns (3) and (iii); the input address level reads now $(B,A)=(1,1)$. Figure \ref{fig:RK} (a-3) displays the real-space dynamics of hot polaritons at 1.5415 eV, propagating rapidly and decaying to lower energy states: it is clearly seen that $A_R$ surpasses the position of the \emph{B} pulse, before the latter impinges on the sample, in its way towards the edge of the ridge. Now, this edge and the double excitonic barrier, composed by $V_A$ and $V_B$, determine the trajectories of the polariton condensates, as shown in Fig.~\ref{fig:RK} (b[t]-3). The creation of $V_B$ is delayed by $\Delta t$, which is precisely set to 80 ps in order to allow the passage of the $A_R$ condensate and its subsequent trapping together with $B_R$, Fig.~\ref{fig:RK} (b-3). The combination of $V_A$ and $V_B$ constitutes a potential trap, in which $B_L$ oscillates periodically. Its amplification by the excitons at \emph{A} and \emph{B} is proven by a significantly stronger emission intensity than that shown in Fig.~\ref{fig:RK} (b-2) for $B_L$, where $V_A$ was absent. The confinement of polaritons in a potential trap increases their density and when polaritons exceed an occupation threshold (at a certain high energy level), they scatter, relaxing into lower energy states.~\cite{Wertz2010,Tosi2012a} This stimulated scattering process mediates the creation of $\mathscr{C}_{B \wedge A}$: it occurs only when $A_R$ and $B_R$ are simultaneously confined between $V_B$ and the ridge's border, resulting in an over-populated energy state at 1.540 eV compared to that shown in Fig.~\ref{fig:RK} (b-2), where only $B_R$ was present. The former situation, shown in Fig.~\ref{fig:RK} (b-3), evolves so that part of the population of the $A_R$ and $B_R$ condensates relaxes and gives rise to the $\mathscr{C}_{B \wedge A}$ condensate, Fig.~\ref{fig:RK} (c-3), which lasts for $\sim$200 ps displaying weak oscillations. \RC{A detailed study of the formation of such a trapped condensate, using different (but equivalent for the trapping process) beam-configuration conditions, has been performed in Ref. \onlinecite{Anton2013prb}.} The $\mathscr{C}_{B \wedge A}$ lifetime, as already shown in Fig.~\ref{fig:setup} (d), is notably larger than that of those condensates expelled far away from the laser position, since the presence of the excitonic reservoir continually feeds $\mathscr{C}_{B \wedge A}$.

The dynamics of the [ON] state in \textbf{k}-space is summarized in the sixth column, (iii), of Fig.~\ref{fig:RK}. In Fig.~\ref{fig:RK} (a-iii), the two populations of hot polaritons, delayed with respect to each other, manifest the same behavior as that obtained for the individual emissions, reported separately in Figs.~\ref{fig:RK} (a-i,ii). Figure \ref{fig:RK} (b-iii) displays a complex distribution of polariton momentum from $t\approx100$~ps to $t\approx400$~ps, since $A_R$, $B_L$, and $B_R$ momenta are superimposed in a range of $\left|k_x\right|\apprle1.5$~$\mu$m$^{-1}$. $A_R$ and $B_R$ suffer elastic collisions almost simultaneously against the ridge's border at $\sim$150 and $\sim$250 ps (dashed segments in (b[t]-iii)); their dynamics is similar to the one of $B_R$ reported in Fig.~\ref{fig:RK} (b[t]-ii). On its own account, $B_L$ shows a zig-zag movement in \textbf{k}-space, since it suffers continuous accelerations/decelerations in the $V_A+V_B$ potential sculpted by the $A$ and \emph{B} beams. In this case, the absence of collisions against a hard-well potential (as described for $A_R$ and $B_R$) yields only slanted trajectories from $\pm 1.0$ to $\mp 1.0$~$\mu$m$^{-1}$ for the $B_L$ movement. The conspicuous interference patterns around $\left|k_x\right|\apprle0.9$~$\mu$m$^{-1}$ arise from the mutual coherence between different condensates. Finally, Fig.~\ref{fig:RK} (c-iii) shows the quasi-steady dynamics of the $\mathscr{C}_{B \wedge A}$ condensate ([ON] state): the confined population sustains an oscillatory movement with a period of 25 ps and an amplitude of $\sim0.75$~$\mu$m$^{-1}$. The $\mathscr{C}_{B \wedge A}$ effective lifetime is determined by the excitons at \emph{B}, which feed the scattering process towards this final state, lasting for more than 200 ps, as reported in Fig.~\ref{fig:setup} (d).

\RD{

\subsection{Unsuitable conditions for \emph{AND} gating: a particular case}
\label{subsec:out}

The choice of the spatial-temporal coordinates of the \emph{A} and \emph{B} beams is crucial to obtain the \emph{AND}-type logic gate. We illustrate this fact in this Subsection presenting a particular case, among the plethora of possible choices for these coordinates, where a trapped condensate is also formed at the ridge's edge but the \emph{AND} operation is not obtained. The spatial distance and the temporal delay between \emph{A} and \emph{B} are $\sim$20 $\mu$m and $\sim$25 ps, respectively. The position of the \emph{A} beam is $45$ $\mu$m away from the right border. Due to the fact that the beams are closer together to each other and also to the edge of the ridge, the used power beam are slightly lower, $P_A= 6\times P_{th}$ and $P_B= 1.5\times P_{th}$,  than those used in Subsec.~\ref{subsec:and}.

Here we present only the polariton dynamics in real space, as summarized in Fig. ~\ref{fig:RK_c}, where again the emission along $y$ has been integrated.  We follow the same nomenclature for the polariton wave trains as the one used before in Fig.~\ref{fig:RK}. Due to the rich dynamics obtained under these new excitation conditions, and its strong dependence on the detection energy, four selected energies (a) 1.5411 eV, (b) 1.5405 eV, (c) 1.5400 eV and (d) 1.5396 eV are now shown. 

\begin{figure}[!htbp]
\setlength{\abovecaptionskip}{-5pt}
\setlength{\belowcaptionskip}{-2pt}
\begin{center}
\includegraphics[trim=0.5cm 0.2cm 0.1cm 0.3cm, clip=true,width=1.0\linewidth,angle=0]{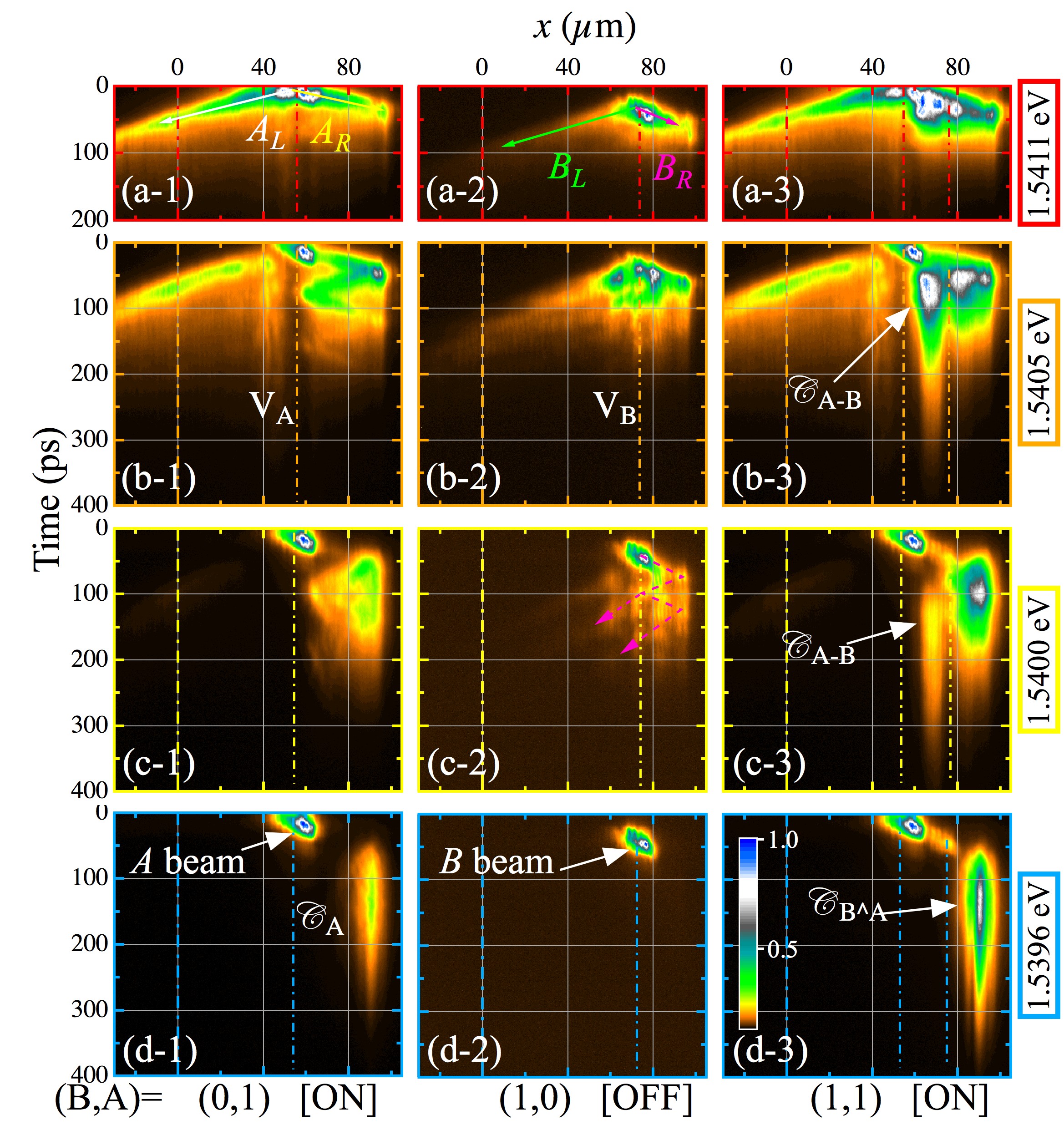}
\end{center}
\caption{\RD{(Color online) Real-space dynamics of the polariton emission, exciting with three logic address inputs, $(B, A)$ (see lower labels): $(0,1)$ \emph{A}-only [column (1)]; $(1,0)$ \emph{B}-only [column (2)] and $(1,1)$ $A+B$ beams [column (3)], at different detection energies: 1.5411 eV (a), 1.5405 eV (b), 1.5400 eV (c) and 1.5396 eV (d). The trajectories of the polariton wave trains $A_{L/R}$ and $B_{L/R}$ at 1.5411 eV are sketched by colored arrows in panels (a-1) and (a-2), respectively. The re-amplification and tunneling of $B_R$  is marked with dashed arrows in panel (c-2) (see text for further details). The intensities are coded in a logarithmic, normalized, false color scale, shown in the panel (d-3). The same spatial-temporal origin is used as that shown in Fig.~\ref{fig:RK}.}}
\label{fig:RK_c}
\end{figure}

The first row (a) in Fig.~\ref{fig:RK_c} shows similar dynamics to those already shown in Subsec.~\ref{subsec:and}, Fig.~\ref{fig:RK}(a), with hot polaritons rapidly decaying into lower energies. Figure~\ref{fig:RK_c} (b-1) reveals several reflections of $A_R$ at the ridge's border ($x=$100 $\mu$m) and at $V_A$ ($x=$55  $\mu$m). At 1.5400 eV, Fig.~\ref{fig:RK_c}(c-1), the zig-zag movement of $A_R$ overlaps with the incipient formation of a trapped condensate, dubbed $\mathscr{C}_{A}$, which is clearly observed at $x=$ 90 $\mu$m and 1.5396 eV, Fig.~\ref{fig:RK_c}(d-1). Therefore, under this single-beam excitation, a trapped condensate is already created, lasting for more than 300 ps, evidencing that these excitation conditions are unsuitable to achieve an \emph{AND} operation.

Column (2) in Fig.~\ref{fig:RK_c} compiles the polariton emission under \emph{B}-only excitation. Figure~\ref{fig:RK_c}(b-2) shows that the lower intensity used in this case for the \emph{B} beam creates a $B_{R}$ wave train with a shorter propagation and weaker emission than $A_{R}$. Figure~\ref{fig:RK_c}(c-2) depicts the $B_R$ re-amplification and its tunneling trough the $V_B$ barrier~\cite{Wertz:2012ee} (see the magenta arrows as a guide to the eye for the tunneling processes). At 1.5396 eV, Fig.~\ref{fig:RK_c}(d-2), there is not any trace of a trapped condensate and only the scattered light by the \emph{B} pulse is observed.

Finally, we show in column (3) of Fig.~\ref{fig:RK_c} the polariton emission under \emph{A+B} excitation. The propagation at high energy, 1.5411 eV, of polaritons rapidly decaying into lower energy states is shown in Fig.~\ref{fig:RK_c}(a-3): in this case, the proximity of the \emph{A} and \emph{B} beams results in an apparent overlap of the $A_R$ and $B_R$ emission, but the choice of $\Delta t\approx 25$ ps still allows that $A_R$ surpasses the barrier $V_B$ created by the \emph{B} beam. The short $A-B$ distance used in this configuration creates a trapped condensate between them, dubbed as $\mathscr{C}_{A-B}$, lasting for more than 200 ps, Fig.~\ref{fig:RK_c}(b-3). The polariton emission at the ridge's edge (region between 80 and 100 $\mu$m) lasts $\lesssim$150 ps at this energy, polaritons rapidly decay in energy forming a trapped condensate, $\mathscr{C}_{B\wedge A}$. Figure~\ref{fig:RK_c}(c-3) shows the coexistence of both $\mathscr{C}_{A-B}$ and $\mathscr{C}_{B\wedge A}$ condensates at 1.5400 eV, the latter being three times more intense than the former one. At 1.5396 eV, Fig.~\ref{fig:RK_c}(d-3) shows the output signal $\mathscr{C}_{B\wedge A}$. This was the [ON] state under suitable excitation conditions (Subsec.~\ref{subsec:and}), which appeared only for the $(B,A)=(1,1)$ input address, however in the present excitation conditions such a signal appears also in the $(B,A)=(0,1)$ case, violating the \emph{AND} truth table. 

It is remarkable that the $\mathscr{C}_{A-B}$ emission, Fig.~\ref{fig:RK_c}(b-3), is 1 meV blue shifted with respect to the $\mathscr{C}_{B\wedge A}$, Fig.~\ref{fig:RK_c}(d-3). This difference is due to the combination of the $V_A$ and $V_B$ barriers and the short distance between them: the potential created by the barriers increases the $\mathscr{C}_{A-B}$ energy, while $\mathscr{C}_{B\wedge A}$ emits from the available lower energy states close to the ridge's border (shown in the inset of Fig.~\ref{fig:disp_rel}(c)). A comparison between the trapped condensates $\mathscr{C}_{A}$ and $\mathscr{C}_{B\wedge A}$, Figs.~\ref{fig:RK_c}(d-1,3), shows that the $\mathscr{C}_{B\wedge A}$ emission is $4$ times stronger and lasts $\sim 50$ ps longer than the $\mathscr{C}_{A}$ emission. The addition of $A_R$ and $B_R$ populations and the re-feeding of $\mathscr{C}_{B\wedge A}$ in presence of the $V_B$ exciton barrier yield this intense and long-living condensate. 

From these experiments we conclude that (i) a single wave train ($A_R$) is able to create a trapped condensate ($\mathscr{C}_{A}$) close to the border (\textbf{unsuitable} condition for the \emph{AND} operation), (ii) inadequate parameters of beam power and distance to the border hinder the formation of this trapped condensate (case of the \emph{B} beam excitation), and (iii) the photo-generated excitonic potentials permit to create blue shifted, trapped condensates on demand (as it has been demonstrated by the creation of $\mathscr{C}_{A-B}$).
}

%%%%%%%%%%%%%%%%%%%%%%%%%%%%%%%%%%%%%%%
%Model
%%%%%%%%%%%%%%%%%%%%%%%%%%%%%%%%%%%%%%%
\section{Model}
\label{sec:model}

\begin{figure}[!htbp]
\setlength{\abovecaptionskip}{-5pt}
\setlength{\belowcaptionskip}{-2pt}
\begin{center}
\includegraphics[trim=0.3cm 0.2cm 0.1cm 0.3cm, clip=true,width=1.0\linewidth,angle=0]{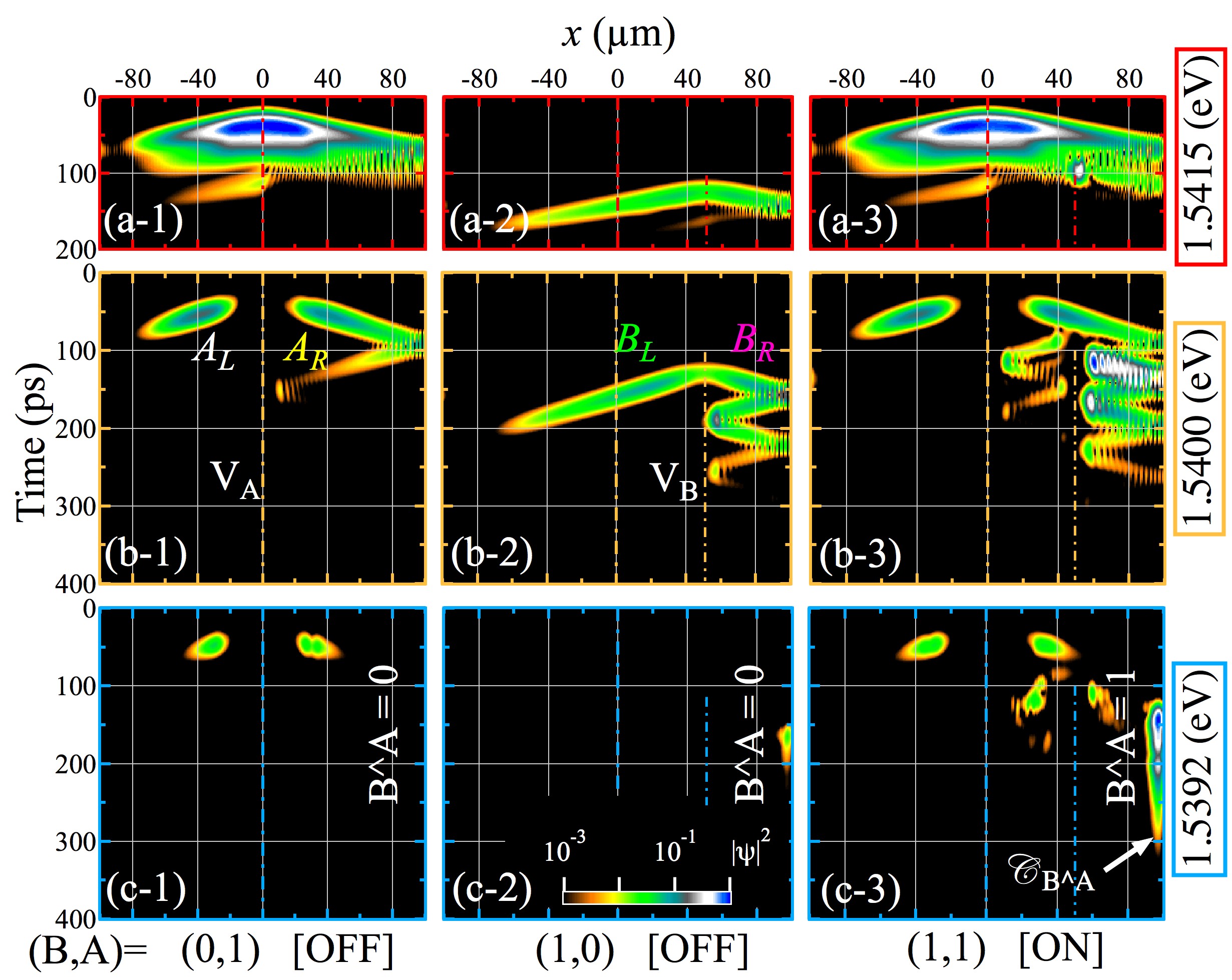}
\end{center}
\caption{\textmoved{(Color online) Calculated real-space dynamics of the polariton emission, filtered at high (a), intermediate (b) and low (c) energies. The columns (1), (2) and (3), correspond to the cases $(B,A)=(0,1)$, $(1,0)$ and $(1,1)$, respectively, as in the experimental Fig.~\ref{fig:RK}. The intensity is coded in a logarithmic, normalized, false color scale, shown in (c-2).}}

\label{fig:model}
\end{figure}

Polariton dynamics can be modeled using a generalized Gross-Pitaevskii description in which evolution equations for the polariton mean-field wavefunction are coupled to a system of semiclassical rate equations for higher energy excitations.~\cite{Wouters2007a} \textmoved{In addition to polariton-polariton scattering, further energy-relaxation processes can be modeled via a phenomenological energy-relaxation term,~\cite{Wouters:2012fk} successfully applied to various microwire experiments.~\cite{Wertz:2012ee,Anton2013prb} The evolution of the polariton wavefunction $\psi(x,t)$ is given by:
\begin{align}
i\hbar\frac{d\psi(x,t)}{dt}&=\left[\hat{E}_{LP}+\alpha|\psi(x,t)|^2+V(x,t)\right.\notag\\
&\hspace{10mm}\left.+i\hbar\left( rN_A(x,t)-\frac{\Gamma}{2}\right)\right]\psi(x,t)\notag\\
&\hspace{5mm}+i\hbar\mathfrak{R}\left[\psi(x,t)\right].\label{eq:GP}
\end{align}
$\hat{E}_{LP}$ represents the kinetic energy dispersion of polaritons. $\alpha$ represents the strength of polariton-polariton interactions. The effective potential $V(x,t)=V_0(x)+\hbar g\left(N_A(x,t)+N_I(x,t)+N_D(x,t)\right)$ describes both the static structural potential of the microwire, $V_0(x)$, and a dynamic contribution from a repulsive potential caused by hot excitons in the system (responsible for the pulse induced barriers). In our sample, there is a local minimum near the wire edge that should be included in $V_0(x)$.

As in other time-dependent studies~\cite{Lagoudakis2011} and our previous work~\cite{Anton2013prb} the non-resonant excitation populates multiple reservoir states. $N_A$ represents the density distribution of an ``active'' hot exciton reservoir with the correct energy and momentum for direct stimulated scattering into the condensate. The non-resonant pumping $P(x,t)$, however, excites different states in an ``inactive'' reservoir, which is coupled to the active reservoir.~\cite{Lagoudakis2011} Long-lived dark excitons described by $N_D$ may also be populated giving a delayed contribution to $V(x,t)$. The exciton density dynamics is given by the rate equations:
\begin{align}
\frac{dN_A(x,t)}{dt}&=-\left(\Gamma_A+r|\psi(x,t)|^2\right)N_A(x,t)+t_RN_I(x,t)\\
\frac{dN_I(x,t)}{dt}&=-\left(\Gamma_I+t_R+t_D\right)N_I(x,t)+P(x,t)\\
\frac{dN_D(x,t)}{dt}&=t_DN_I(x,t)-\Gamma_DN_D(x,t)\label{eq:ND}
\end{align}
$\Gamma_A$, $\Gamma_I$ and $\Gamma_D$ describe the decay rates of each of the exciton types. $t_R$ and $t_D$ are inter-reservoir coupling constants, while $r$ gives the rate of condensation, which enters into Eq.~\ref{eq:GP}. $\Gamma$ is the decay rate of polaritons.

The final term in Eq.~\ref{eq:GP} accounts for energy relaxation processes of condensed polaritons:
\begin{equation}
\mathfrak{R}[\psi(x,t)]=-\left(\nu+\nu^\prime|\psi(x,t)|^2\right)\left(\hat{E}_\mathrm{LP}-\mu(x,t)\right)\psi(x,t),\label{eq:relax}
\end{equation}
where $\nu$ and $\nu^\prime$ are phenomenological parameters determining the strength of energy relaxation~\cite{Wouters:2012fk,Wertz:2012ee} and $\mu(x,t)$ is a local effective chemical potential that conserves the polariton population. More details on these terms can be found in Refs.~\onlinecite{Anton2013prb} and \onlinecite{Wouters:2012fk}. The terms cause the relaxation of any kinetic energy of polaritons and allow the population of lower-energy states trapped between the pump-induced potentials.

Equations~\ref{eq:GP}-\ref{eq:ND} were solved numerically with Gaussian pulses corresponding to the experimental configurations, with the results shown in Fig. \ref{fig:model}.~\cite{andprlparam} The pulses, described by $P(x,t)$, had the same shape, size and duration but were separated by $80$~ps and $50$~$\mu$m. The $B$ pulse was taken with $60$\% the intensity of the $A$ pulse, similar to the experimental situation discussed in Subsec.~\ref{subsec:and}.}

At high energies, the pulses create polariton condensates at their impinging positions, which spread out over time. After a time delay, the polaritons relax their energy entering trapped states, which undergo multiple reflections depending on the excitation configuration, as in the experimental case. In the case when both pulses are present, $(B,A)=(1,1)$, an enhanced polariton density in a trapped state near the wire edge allows further stimulated energy relaxation, populating the low energy condensate, $\mathscr{C}_{B \wedge A}$, seen in Fig.~\ref{fig:model}(c-3). The localization of this condensate is partly due to a static potential $V_0(x)$, which is known to have a minimum near the wire edge. Note that the theory only reproduces the emission from condensed polaritons; it does not show the scattered light responsible for the peaks at the pulse positions and arrival times seen in Fig.~\ref{fig:RK}.

%%%%%%%%%%%%%%%%%%%%%%%%%%%%%%%%%%%%%%%
%Conclusions
%%%%%%%%%%%%%%%%%%%%%%%%%%%%%%%%%%%%%%%
\section{Conclusions}
\label{sec:conclusions}

Using time- and energy-resolved measurements, we studied the dynamics of polariton condensate wave trains propagating in a 1D semiconductor microcavity ridge. The introduction of dynamic, fully controllable potential barriers allows the reflecting and re-routing of the wave trains. Multiple reflections occurring between pairs of barriers (optical induced or formed by the wire edge) allow the confinement and storage of propagating polariton states. This localization also gives rise to an enhancement of energy relaxation into a coherent ground-state, which forms a convenient output of a logical \emph{AND} gate. The total dynamics in real- and momentum-space has been tracked yielding the modus operandi of the device. Theoretical simulations based on a generalized Gross-Pitaevskii description reproduce the experimental dynamics and confirm our understanding of the system. These results pave the way for the realization of ultrafast, compact switches based on the manipulation of Bose-Einstein-like condensates. Due to the use of incoherent/non-resonant excitation, this framework provides a path toward hybrid electro-optical information processing devices and offers additional functionalities for different applications ranging from coherent matter transport to interferometry.

%%%%%%%%%%%%%%%%%%%%%%%%%%%%%%%%%%%%%%%
%Acknowledgments
%%%%%%%%%%%%%%%%%%%%%%%%%%%%%%%%%%%%%%%
\section{Acknowledgments}
\label{sec:acknow}

We thank G. Tosi for fruitful discussion. \textmoved{C.A. and J.C. acknowledge financial support from a Spanish FPU scholarship and Marie Curie Project, respectively}.  P.S. acknowledges Greek GSRT program ``ARISTEIA'' (1978) and EU ERC ``Polaflow'' for financial support. Partial support was also obtained from the Spanish MEC MAT2011-22997, CAM (S-2009/ESP-1503) and FP7 ITN's ``Clermont4'' (235114), ``Spin-optronics" (237252) and INDEX (289968) projects.

%%%%%%%%%%%%%%%%%%%%%%%%%%%%%%%%%%%%%%%
%Appendix
%%%%%%%%%%%%%%%%%%%%%%%%%%%%%%%%%%%%%%%
\textmoved{

\section*{Appendix A: List of symbols}
\label{sec:symbols}

In this Appendix we define the symbols used as abbreviations along the manuscript.

\begin{table}[h]
\centering
\begin{tabular*}{1\columnwidth}{@{\extracolsep{\fill}}|c|>{\centering}p{0.8\columnwidth}|}
\hline
\bf{Symbol} & \bf{Meaning}\tabularnewline
\hline
\hline
$\Omega_R$ & Rabi splitting\tabularnewline
\hline 
$E_X$ & Bare exciton mode\tabularnewline
\hline
$E_C$ & Bare cavity mode\tabularnewline
\hline 
$\delta$ & Cavity detunning defined as $\delta=E_C-E_X$ \tabularnewline
\hline
$A$ & First beam impinging at the ridge at certain $\{x,t\}$ coordinates\tabularnewline
\hline
$B$ & Second beam impinging at the ridge at certain $\{x,t\}$ coordinates\tabularnewline
\hline
$\Delta t$ & Relative time delay $t_B-t_A$ between \emph{A} and \emph{B} beams\tabularnewline
\hline
$P_{th}$ & Power threshold to produce polariton condensates\tabularnewline
\hline
$P_A$ & \emph{A} beam power\tabularnewline
\hline
$P_B$ & \emph{B} beam power\tabularnewline
\hline
$V_A$ & Photo-generated excitonic potential created by the \emph{A} beam \tabularnewline
\hline
$V_B$ & Photo-generated excitonic potential created by the \emph{B} beam\tabularnewline
\hline
$\mathscr{C}_{B\wedge A}$ & Trapped condensate between $V_B$ and the right ridge's border\tabularnewline
\hline
$A_{L/R}$ & Initially propagating left/right polariton wave trains created by the \emph{A} beam\tabularnewline
\hline
$B_{L/R}$ & Initially propagating left/right polariton wave trains created by the \emph{B} beam\tabularnewline
\hline
$\mathscr{C}_{A}$ & Trapped condensate between $V_A$ and the right ridge's border created under only \emph{A} beam excitation\tabularnewline
\hline
$\mathscr{C}_{A-B}$ & Trapped condensate between $V_A$ and $V_B$ under adequate excitation conditions\tabularnewline
\hline

\end{tabular*}

\caption{List of symbols used on Sec.~\ref{sec:exp} Experimental results and discussion.}
\label{tablexp}
\end{table}

%%%%%%%

\begin{table}[h]
\centering
\begin{tabular*}{1\columnwidth}{@{\extracolsep{\fill}}|c|>{\centering}p{0.8\columnwidth}|}
\hline
\bf{Symbol} & \bf{Meaning}\tabularnewline
\hline
\hline
$\psi(x,t)$ & Polariton wavefunction\tabularnewline
\hline
$\hbar$ & Planck constant divided by 2$\pi$\tabularnewline
\hline
$E_{LP}$ & Lower polariton branch energy dispersion\tabularnewline
\hline
$\alpha$ & Polariton-polariton interaction strength\tabularnewline
\hline
$V\left(x,t\right)$ & Effective polariton potential\tabularnewline
\hline
$r$ & Polariton condensation rate\tabularnewline
\hline
$N_A$ & Density distribution of active excitons\tabularnewline
\hline
$\Gamma$ & Polariton decay rate\tabularnewline
\hline
$\mathfrak{R}$ & Energy relaxation term of condensed polaritons\tabularnewline
\hline
$V_0\left(x\right)$ & Wire structural potential\tabularnewline
\hline
$g$ & Polariton-exciton interaction strength \tabularnewline
\hline
$N_I$ & Density distribution of inactive excitons\tabularnewline
\hline
$N_D$ & Density distribution of dark excitons\tabularnewline
\hline
$P(x,t)$ & Non-resonant pumping\tabularnewline
\hline
$\Gamma_A$ & Decay rate of the active exciton reservoir\tabularnewline
\hline
$\Gamma_I$ & Decay rate of the inactive exciton reservoir\tabularnewline
\hline
$\Gamma_D$ & Decay rate of the dark exciton reservoir\tabularnewline
\hline
$t_R$ & Linear coupling to the inactive exciton reservoir\tabularnewline
\hline
$t_D$ & Linear coupling to the dark exciton reservoir\tabularnewline
\hline
$\nu$/$\nu^\prime$ & Phenomenological parameters for the strength of energy relaxation\tabularnewline
\hline
$\mu(x,t)$ & Local effective chemical potential\tabularnewline
\hline
$m$ & Polariton effective mass\tabularnewline
\hline
$m_e$ & Free electron mass\tabularnewline
\hline
\end{tabular*}

\caption{List of symbols used on Sec.~\ref{sec:model} Model.}

\label{tableteo}

\end{table}

}
%%%%%%%%%%%%%%%%%%%%%%%%%%%%%%%%%%%%%%%
%Bibliography
%%%%%%%%%%%%%%%%%%%%%%%%%%%%%%%%%%%%%%%

%\bibliographystyle{wihuri}
%\bibliography{Biblio1}

\end{document}